# Towards Data-driven GIM tools: Two Prototypes

**Dinneen, Jesse David**    Humboldt-Universität zu Berlin, Germany
                            jesse.dinneen@hu-berlin.de
**Donner, Sascha**          evana.ai, Germany
**Bubinger, Helen**         Humboldt-Universität zu Berlin, Germany
**Low, Jwen Fai**           McGill University, Canada
**Krtalić, Maja**           Victoria University of Wellington, New Zealand

**ABSTRACT**

Here we describe two approaches to improve group information management (GIM) and draw on the results of prior works to implement them in software prototypes. The first aids browsing and retrieving from large and unfamiliar collections like shared drives by dynamically reducing and re-organising them. The second supports the transfer and re-use of collections (e.g. to/by successors, descendants, or curators) by integrating novel sorting and annotation features. The prototypes' source code is shared online and screenshots are presented in the accompanying poster.

**INTRODUCTION**

People in the information society are tasked everyday with storing, organising, and refinding an increasing number and variety of digital items in a challenging task known as *personal information management* (PIM; Jones et al., 2017). Such activities are made further challenging when other people are involved in some way – in *group information management* (GIM) – for example when finding items in collections organised by coworkers or when organising a collection that will be later re-used by family (Dinneen & Julien, 2020). To help address the challenges of GIM we offer two prototype interfaces, the ideation and design of which are informed by data collected in prior studies. Both prototypes were developed using the Python programming language and Qt graphics framework. *Their interfaces can be seen on the accompanying poster and their source code is shared online* (links below).

**GIMZOOMER**

Personal and group-managed collections of digital files are often very large, consisting of tens of thousands of files and folders organised into tree shapes that are deep and wide (Dinneen et al., 2019). Large information structures of all kinds are usually difficult to browse (Julien et al., 2013) and therefore also to organise and retrieve from; for example they require users to review and decide among many labels (e.g. names of folders, genres, nodes) and maintain a sense of their location as they navigate down and across the tree and identify one or more relevant items (Savolainen, 2018). Group-managed structures introduce further challenges, as users often lack the benefit of familiarity that comes from having previously seen, organised, or labelled the folders they are browsing or retrieving from (Bergman et al., 2014). As a result, retrievals require more effort, take longer, and are more prone to failure (Bergman et al., 2019, Julien et al., 2013). In some cases, users may customise the structure to suit their individual retrieval strategy and aid later relevance judgments (de Fremery & Buckland, 2022), but doing so may violate group conventions or decrease the intelligibility of the structure for others users (Berlin et al., 1993; Dourish et al., 1999; Erickson, 2006). Such problems manifest today on shared drives during collaboration, turnover, and onboarding.

One way to address large and unfamiliar collections is to provide flexible views wherein users can see and modify the collection in a way that benefits them and is determined less by other users (Dourish et al., 1999; Ellis & Dix, 2007). A promising start to this approach is to reduce the structure so that the displayed portion and its overall browsing complexity are reduced, as has been done with two information structures that organise published scientific texts: LCSH and MeSH (Julien et al., 2013). There, the extremely uneven distribution of most books into a relatively small number of genres enabled two algorithmic reductions: (a) compressing the structure by moving book-heavy genres (and their siblings) up in the tree, and (b) pruning (i.e. removing) book-light genres. With this approach the browsing complexity of MeSH was reduced by ~80%, and of LCSH by ~50% without precluding access to the majority of the books. Later, a between-subjects experiment confirmed the usefulness of the approach in practice: when using a strongly pruned-and-compressed LCSH tree, participants completed retrieval tasks significantly more accurately and faster, and with no fewer (potentially valuable) interactions with the structure than participants using unmodified or only compressed trees (Dinneen et al., 2018).

Although developed for LCSH and MeSH, the reduction approach may also work well for GIM structures: recent works have established that digital files are often distributed unevenly into folders similarly to the books and genres of LCSH and MeSH (Dinneen et al., 2019), so if a user is retrieving a file, it is likely (*ceteris paribus*) to be located in one of only a few large folders, and rapidly reducing the display to only those folders should enable faster access.

Driven by these observations, we designed GIMZoomer, an interface implementing the compression and pruning algorithms provided by Julien et al. (2013). Specifically, the interface features a slider control between tree views showing (1) an unmodified folder hierarchy and (2) its compressed and pruned version. The slider allows the user to control the strength of the algorithms and thus flexibly and iteratively reduce/increase how many folders are displayed in the second tree view and the maximum downward navigation steps; the *ends* of the slider thus represent extremes of the algorithm strengths, producing no modification (bottom) and the strongest modification (top), wherein only a single folder is shown (the fullest). The user can thus reduce the structure, navigate within it to their desired file or folder, or in more complex cases of information seeking (e.g. berrypicking), continue modifying and exploring the structure. The source code is available at https://github.com/jddinneen/GIMZoomer.

**COLLECTION ANNOTATION ASSISTANT**

We note above that individuals and institutions accumulate digital file collections of extensive scale and variety. For individuals, collections contain a variety of everyday items such as private photos, family plans and finances, professional activities, and information reflecting many other of life's activities and tasks (Dinneen & Julien, 2019; Jones et al., 2017; Krtalić & Ihejirika, 2022). Accumulated over years or a lifetime, these digital assets can have long-term value for family members, and the collections of culturally significant people can also have value for society in the form of cultural heritage (Krtalić et al., 2021). But such collections also pose management challenges that entail various risks of loss (Digital Preservation Coalition, 2021; Krtalić & Dinneen, 2022). When collections are transferred to heirs or cultural heritage repositories (Day & Krtalić, 2021), whether for immediate use or preservation, they are often shared without detailed metadata about the contents or context, and it can thus be very difficult for the new owner or custodian to navigate and grasp the often vast and complex collections (Dinneen & Krtalić, 2020; Society of American Archivists, 2013). There is thus a frequent risk that collections will not be available for current or future generations because parts of the collection are inadvertently deleted, the meaning of the content is not sufficiently clear, too large a portion of the collection is irrelevant (i.e. precludes finding the relevant items), or some portion may be too private to be shared (e.g. insufficient time and labour to filter such items before the collection is shared).

Software developers have acknowledged that such collections or *digital legacies* need to be considered before problems arise. For example, Apple recently released a Legacy Contacts feature where up to five people can be named as managers of one's digital assets after death (Digital Legacy Association, 2016). This has the potential to help collections avoid total loss after death and thus avoid losing their value (in the sense indicated above). There remains, however, no tool available with which to indicate which *portions* of a collection might be relevant to families, curators, etc., nor which parts should not be transferred or seen, and so the practical challenges of identifying, re-using, or excluding relevant portions of a collection remain. While prior works have emphasised the importance of supporting digital legacies (e.g. Dinneen et al., 2016; Jones et al., 2016), no tool has yet been developed to help people make sense of large inherited collections, determine which portions are relevant for specific stakeholders, and ensure that collections can continue to be used in the spirit of the collection creator.

A simple but promising approach to avoiding the problems above is to facilitate annotation of the majority of a collection. We thus designed a Collection Annotation Assistant, a backup-like software where parts of a collection can be marked as relevant or valuable to some context (e.g. to a theme, career, family, institution, or society). In the interface a tree-view presents the folder collection together with tick-boxes to indicate relevancy or, alternatively, exclusion (i.e. unable to be labelled relevant). As collections are often so large that reviewing the relevant portions could be time intensive, we also leverage (a) the uneven collection distribution described above and (b) trends in access times (most folders have not been accessed in the last six months; Dinneen, 2018) to add two related sorting options: by accessible files (i.e. those contained in the folder and sub-folders) and by last date modified. In this way, users should be able to quickly sort the collection and annotate the majority of it in relatively few clicks. A further screen in the interface allows users to note any software required for accessing content stored within files, for example in the case of atypical or obsolete formats. The annotations and notes can then be saved in the interoperable JSON format and loaded later for review or updating. The same interface can thus be used later by the inheritors to get a quick impression of what is in the collection, where the relevant parts are, and what should not be accessed or shared. The source code is available at https://github.com/jddinneen/collection-annotation-assistant.

**CONCLUSION**

Although the prototypes described here should be evaluated (e.g. for their efficacy and usability), demonstrating new approaches to GIM is important step forward because inadequate tools can prevent individuals from acclimating to new work places, finding information needed in daily life (Jones et al., 2017), maintaining and sharing their collections, and preserving them for future generations of family or society (Krtalić et al., 2021).

**ACKNOWLEDGMENTS**

The authors thank Charles-Antoine Julien for helping to create GIMZoomer and creating the related algorithms.

# Towards Data-Driven GIM Tools: Two Prototypes

Jesse Dinneen[1], Sascha Donner[1], Helen Bubinger[1]
Jwen Fai Low[2], Maja Krtalić[3]
1 Humboldt-Uni. zu Berlin, DE
2 McGill Uni., CA
3 Victoria Uni. Wellington, NZ

## Problem

Large, unfamiliar collections are difficult to navigate or retrieve from, and their contents are unusable without context like relevance.

## Approach

Design algorithms to exploit the uneven distribution of collections seen in prior work

Implement features into tools to aid navigation & retrieval, enable adding & reviewing context

Share source code: https://github.com/jddinneen/

## Conclusion

The features function as intended and the tools show promise for helping with GIM tasks. Next the tools should be tested with users in real cases of retrieval and collection transfer.

## GIMZoomer

To navigate and retrieve shared folders **faster**, the user moves the slider (1) up, which increases folder *compression* and *pruning*, producing a smaller tree (2).

## Collection Annotator

**To facilitate the transfer and reuse of a collection**, the user sorts (1) by accessible files to quickly find the majority of the collection, and adds metadata for relevance (2) or exclusion (3) in the checkboxes.

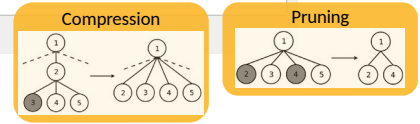

ASIS&T 85TH ANNUAL MEETING